\documentclass[aps,amssymb,prl, amsmath, twocolumn, floatfix]{revtex4}
\usepackage{graphics}
\usepackage{psfrag}
\usepackage{epsfig}
\usepackage{subfigure}
\begin{document}
\bibliographystyle{prsty}

\title{A percolation model for slow dynamics in glass-forming materials}
\author{Gregg Lois}
\author{Jerzy Blawzdziewicz}
\author{Corey S. O'Hern}
\affiliation{
Department of Physics, Department of Mechanical Engineering, Yale
University, New Haven, Connecticut 06520-8284 }

\begin{abstract}
We identify a link between the glass transition and percolation of
mobile regions in configuration space.  We find that many hallmarks of
glassy dynamics, for example stretched-exponential response functions
and a diverging structural relaxation time, are consequences of the
critical properties of mean-field percolation.  Specific predictions
of the percolation model include the range of possible stretching
exponents $1/3 \leq \beta \leq 1$ and the functional dependence of the
structural relaxation time $\tau_\alpha$ and exponent $\beta$ on
temperature, density, and wave number.
\end{abstract}
\maketitle

As temperature is decreased near the glass transition, the structural
relaxation time in glassy materials increases by many orders of
magnitude with only subtle changes in static
correlations~\cite{glassreview}.  In addition, structural correlations
display an anomalous stretched-exponential time decay
$\exp(-t/\tau_\alpha)^\beta$, where $\beta$ is the stretching exponent and
$\tau_\alpha$ the relaxation time.  Understanding the origin of these
behaviors is one of the most important outstanding problems in statistical
physics.

Although stretched-exponential relaxation is common to many
glass-forming materials, the dependence of $\tau_\alpha$ and $\beta$
on temperature and density is not universal.  For molecular, colloidal
and polymer glasses, where structural relaxation is measured using
density autocorrelation functions, the temperature dependence of
$\tau_\alpha$ is affected by the fragility~\cite{angellfragility}.  In
magnetic glasses, where structural relaxation is measured using spin
autocorrelation functions, $\tau_\alpha$ depends on details of the
microscopic interactions~\cite{spinglasstimes}.  In all
glassy systems the stretching exponent $\beta$ varies between
$1/3$ and $1$ depending on scattering wave vector, density and temperature,
and its dependence on these variables is not universal~\cite{ngai}.

How do we understand anomalous relaxation in glass-forming materials
where, although correlation functions display stretched-exponential
relaxation, the temperature and density dependence of $\tau_\alpha$
and $\beta$ vary from one material to the next?  To answer this
question, it is important to consider how the energy landscape, which
is the energy hypersurface as a function of all configurational
degrees of freedom, affects dynamics~\cite{stillingerhiddenstructure}.
Activation out of energy minima is rare at low temperature, and the
system is confined to basins surrounding energy minima with 
infrequent hopping between minima~\cite{heuer} that allows the system
to explore configuration space~\cite{ericweeks}.

In this work, we focus on the connection between anomalously slow
dynamics in glass-forming materials and percolation of mobile domains
in configuration space.  The decay of structural correlations over a
time $t$ is related to the average distance that the system
moves in configuration space during that time.  Thus complete
relaxation---decay of structural correlations to zero---only occurs
after the system diffuses through a network of basins that percolates
configuration space.  For both hard-sphere systems with infinite
energy barriers and systems with finite barriers, we demonstrate that
a percolation transition in configuration space is responsible for
several hallmarks of glassy dynamics: (1) stretched-exponential
relaxation of structural correlations and the experimentally observed
range of values and wave-number dependence of the stretching exponent
$\beta$; (2) the form of the divergence of $\tau_\alpha$, and (3) a
diverging length scale near the glass transition for hard spheres.

{\bf A. Hard spheres:} We first consider collections of hard spheres
that interact at contact with an infinite repulsion.  At moderate
density, hard spheres behave as simple fluids.  However, as density
increases structural relaxation becomes anomalously slow.  Upon
further compression, if crystallization is avoided, the system becomes
confined to a disordered {\em collectively
jammed}~\cite{torquatojamminghardspheres} (CJ) state at packing
fraction $\phi_J$.  We focus on situations where, due to particle
size polydispersity or fast quench rates, crystallization is avoided,
and the systems are disordered. In CJ states at $\phi_J$ all single
and collective particle displacements cause particle overlap; thus,
due to hard-sphere constraints, no motion occurs at $\phi_J$.  A
system of $N$ hard spheres in $d$ dimensions has $\sim N! e^{aN}$ CJ
states~\cite{xu_counting,footnote10} that can be represented as points
in $dN$-dimensional configuration space, where $a>0$ is a constant.

The transition from glass to liquid can be understood as the
percolation of a network of bonds between points that represent CJ
states in configuration space, where a bond is formed when the system
can move between two CJ states without violating hard-sphere
constraints.  At high packing fraction there is no percolating cluster
and relaxation is limited by the maximum cluster size.  At low packing
fraction, a percolating network of bonded CJ states spans
configuration space and the system can fully relax.  To use the
predictions of continuum percolation it is useful to partition
configuration space into {\em basins of attraction}, each associated
with an individual CJ state~\cite{footnote1}. At a given packing
fraction $\phi$ we define a {\em mobility domain} for each CJ state as
the portion of the basin of attraction that satisfies the hard-sphere
constraints at $\phi$.  Thus at $\phi=0$ (when there are no
hard-sphere constraints) the mobility domain and basin of a CJ state
are the same, for $\phi>0$ the mobility domain is a subset of the
basin, and at $\phi_J$ the mobility domain is the CJ state.

\begin{figure}
\scalebox{0.75}{\includegraphics{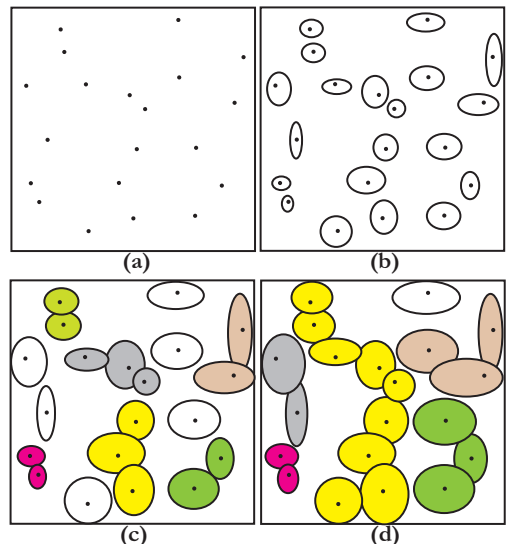}}
\vspace{-0.15in}
\caption{ \label{diagram} (Color online).  Accessible
regions in configuration space for hard spheres.  (a) At
$\phi_J$ only CJ states (points) are accessible; (b) At $\phi_1 <
\phi_J$ motion occurs in mobility domains surrounding CJ
states; (c) At $\phi_2<\phi_1$ transitions between CJ
states with contacting mobility domains (shaded) occur; (d) At $\phi_P <
\phi_2$ at least one mobility domain percolates (shaded yellow) and
the system transitions from glass to liquid.  }
\vspace{-0.15in}
\end{figure}

We formulate a percolation model by considering the motion of hard
spheres near $\phi_J$.  At $\phi_J$ the system is confined to one CJ
state and no motion occurs (Fig.~\ref{diagram}(a)).  For $\phi
\lesssim \phi_J$ there is a closed mobility domain accessible to each
CJ state~\cite{torquatojamminghardspheres}
(Fig.~\ref{diagram}(b)). For smaller $\phi$ the mobility domains of
two CJ states can come into contact, which allows the system to
transition between CJ states (Fig.~\ref{diagram}(c)).  
The system will
diffuse on a network of mobility domains if many are in contact. We
also expect that when the mobility domains become sufficiently large,
a percolating cluster of domains is formed (Fig.~\ref{diagram}(d)).

Structural relaxation in dense hard-sphere systems occurs via
dynamical heterogeneities~\cite{coopmotion}, which implies that the
shape of mobility domains is complex.  Near CJ states mobility domains
are roughly $dN$-dimensional and are quickly explored.
Further from CJ states, the effective dimension of mobility domains
can be significantly reduced, and the time needed to explore these
regions is large.  We also expect polydispersity of mobility domain
sizes for different CJ states.  However, we assume there is an upper
critical dimension $D^*$ of configuration space above which mean-field
theory accurately describes critical exponents.  In this limit,
complexities arising from the geometry and correlations of mobility
domains can be ignored.  We focus on large $N$ where $dN>D^*$ and
construct a mean-field theory in terms of the packing fraction of
mobility domains $\Pi$ in configuration space.  Percolation occurs at
a critical value $\Pi_P$ and is controlled by the mean-field
percolation exponents~\cite{staufferpercolationbook}.  We employ the
mean-field assumption throughout this section, and outline a
non-mean-field approach at the end of section~B.

Our percolation model predicts a glass for $\Pi < \Pi_P$ and
liquid for $\Pi > \Pi_P$.  Structural relaxation of hard spheres
can be quantified using the incoherent part of the intermediate
scattering function (ISF) $\Phi_{\vec{q}}(t) = \sum_j \exp{[i \vec{q}
\cdot \Delta \vec{r}_j(t)]}/N$, where $\Delta \vec{r}_j(t)$ is the
displacement of particle $j$ over time $t$ and $\vec{q}$ is
the scattering wave vector~\cite{ISF}.  The infinite-time value of the
ISF $f_{\vec{q}} \equiv \Phi_{\vec{q}}(\infty)$ is an order parameter
for the glass transition that is zero for a liquid and positive for a
glass.  For $\Pi < \Pi_P$ there is no percolating cluster of mobility
domains, $f_{\vec{q}}>0$, and the system is a glass since the maximum
distance it can diffuse, which is set by the percolation correlation
length $\xi \propto (\Pi_P-\Pi)^{-1/2}$, is finite.  Using a
cumulant expansion we predict $\log{f_{\vec{q}} \propto -q^2 \xi^2
\propto -q^2 (\Pi_P-\Pi)^{-1}}$ for $q \xi \ll 1$.  For $\Pi > \Pi_P$
the largest cluster percolates and $f_{\vec{q}}=0$.

Time dependence of the ISF can be determined from  
dynamics in configuration space.  The model contains two important
time scales: short times where the system is confined to a
single mobility domain and long times where the system
diffuses on a connected cluster of mobility domains.  The short-time
dynamics is characterized by the time $\tau_0$ needed to transition
from one mobility domain to another.  When $\Pi>\Pi_P$ and
a percolating network exists, long-time relaxation of the ISF is
characterized by anomalous diffusion on fractal percolating networks.
The mean-square displacement (MSD) $\Delta r^2(t)$ on percolating networks
obeys the scaling form
\begin{equation}
\Delta r^2(t) \propto (t/\tau_0)^{1/3} G\left[\xi^{-2} (t/\tau_0)^{1/3}\right],
\label{diffusionr2}
\end{equation}
where $\xi\propto (\Pi-\Pi_P)^{-1/2}$ is the percolation correlation
length and the scaling function $G[z] \propto z^2$ for $z \gg 1$ and
$G[z]=1$ for $z \ll 1$~\cite{percdiffusion,percdiffusion2}.  To first
order in a cumulant expansion $\Phi_{\vec{q}}(t) = \exp(-q^2 \Delta
r^2(t)/6)$, and the alpha relaxation time $\tau_\alpha$, defined via
$\Phi_{\vec{q}}(\tau_\alpha) = e^{-1}$, is \begin{equation}
\tau_\alpha \propto \Bigg\{ \begin{array}{cc} \tau_0 \, q^{-2}
(\Pi-\Pi_P)^{-2} & \mathrm{for} \, \, q\xi \ll 1, \\ \tau_0 \, q^{-6}
& \mathrm{for} \, \, q\xi \gg 1.
\end{array}
\label{relaxtime}
\end{equation}
Note that there are two contributions to $\tau_\alpha$: (1) the
average transition time between mobility domains $\tau_0$ and (2) a
time-scale for diffusion on percolating networks, which results in the
factor $(\Pi-\Pi_P)^{-2}$.  The $q$-dependence in
Eq.~(\ref{relaxtime}) is consistent with experiments on dense
colloidal suspensions~\cite{vanMegen} and vibrated granular
materials~\cite{PReis}.  Since the MSD in Eq.~(\ref{diffusionr2})
crosses over from anomalous diffusion $\Delta r^2(t) \propto t^{1/3}$
at small times to normal diffusion $\Delta r^2(t) \propto t$ at large
times, the percolation model predicts that $\beta$ varies with time
and satisfies $1/3 \leq \beta \leq 1$.  In experiments, $\beta$ is
measured by fitting the ISF to $\exp(t/\tau_\alpha)^\beta$ near
$\tau_\alpha$, and thus time dependence has not been observed.  Near
$\tau_\alpha$ the percolation model predicts that $\beta$ only depends
on $q \xi$ with $\beta = 1/3$ for $q \xi \gg 1$ and $\beta = 1$ for $q
\xi \ll 1$.  These limiting values for large and small $q$ have been
observed in experiments of hard spheres~\cite{PReis}.

To further test the predictions of the percolation model we express
$\Pi$ and $\tau_0$ in terms of $\phi$.  $\Pi$ is equal to the number
of CJ states ($N! e^{aN}$) multiplied by the average volume $V$ of
each mobility domain in configuration space (normalized by the total
volume $L^{Nd})$.  $V$ is determined by the total free volume $v_f$
accessible to the system in real space and to first order $V^{1/N} =
v_f/(L^d) = (\phi_J-\phi)/\phi_J$.  Thus for large $N$, $\Pi = N^N
(1-\phi/\phi_J)^N$.  In the mean-field approximation, $\tau_0$ is
inversely proportional to the average number of mobility domains in
contact with any single domain.  Assuming a random distribution of
hyperspherical mobility domains, we derive $\tau_0 \propto 2^{dN}
\Pi$.

The packing fraction $\phi_P$ where percolation of mobility domains
occurs is $\Pi_P = N^N (1-\phi_P/\phi_J)^N$.  Mean-field
percolation in configuration space occurs when the fraction of
contacting mobility domains is $1/(z-1)$, where $z$ is the average
number of mobility domains that contact any single one in the $\phi
\rightarrow 0$ limit~\cite{staufferpercolationbook}.  Since at
percolation the average number of mobility domains that contact any
single domain is $2^{dN} \Pi_P$, then $2^{dN} \Pi_P/z = 1/(z-1)$.
Since $z$ increases with $N$, we predict $(\phi_J-\phi_P) \propto
N^{-1}$ for large $N$.  Thus, $\phi_P < \phi_J$ for finite
$N$ whereas $\phi_P=\phi_J$ as $N \rightarrow \infty$.  The
dependence of $\phi_P$ on system size $N$ suggests a diverging length
scale $\ell$ in real space.  For a system of hard spheres at packing
fraction $\phi$, relaxation can only occur in subsystems of size
$\mathcal{N} \sim \ell^d$ where $\phi_P(\mathcal{N}) > \phi$.  Using
$(\phi_J-\phi) \propto 1/\mathcal{N}$ we predict $\ell \propto
(\phi_J-\phi)^{-1/d}$.

We can now express $\tau_\alpha$ in terms of $\phi$.  Since 
$\tau_0 \propto 2^{-dN} \Pi^{-1} \propto
\exp[A\, \phi_J/(\phi_J-\phi)]$ for large
$N$~\cite{footnote3}, the packing fraction dependence of $\tau_\alpha$
can be determined from Eq.~(\ref{relaxtime}).  For $q \xi \ll 1$,
\begin{equation}
q^{2} \tau_\alpha \propto \Bigg\{ \begin{array}{ll} 
\vspace{.05 in}
\exp[\frac{A\, \phi_J}{\phi_J-\phi}]\, (\phi_P-\phi)^{-2} & \mathrm{for} \, \, \phi_P-\phi \ll \phi_J-\phi_P \\
\exp[\frac{B\, \phi_J}{\phi_J-\phi}] & \mathrm{for} \, \, \phi_P-\phi \gg \phi_J-\phi_P,
\end{array}
\end{equation}
where $A$ and $B$ are positive constants.  In the large-$N$ limit when
$\phi_P=\phi_J$ the model predicts a Vogel-Fulcher divergence at
$\phi_J$.  In this limit, the functional form and location of the
divergence have been verified in experiments of hard
spheres~\cite{PReis, hsexponential}.  For finite $N$ when
$\phi_P<\phi_J$ there is a power-law divergence for $\phi$ near
$\phi_P$ and Vogel-Fulcher behavior far from $\phi_P$.

Finally, we emphasize that there are key differences between
disordered hard-sphere systems in- and out-of-equilibrium.  The
largest connected cluster of mobility domains has greater entropy
than all other clusters, and equilibrium relaxation is described by the
arguments above.  However for large quench rates, the system 
is confined to a network of mobility domains that is detached from
the percolating cluster.  This results in a non-equilibrium glass
state with $\phi<\phi_P$, where $f_{\vec{q}}$ is set by the size of 
non-percolating clusters, and the location of the glass transition
depends on system preparation.

{\bf B. Finite energy barriers:} In contrast to hard spheres,
activation is important in systems with finite energy barriers.  
We now extend the percolation model to include activated processes in
systems at constant temperature $T$ (with $k_B=1$).  We again assume
that only disordered states exist, as in frustrated geometries
such as the pyrochlore lattice, polydisperse colloidal suspensions,
and metallic glasses above the critical quench rate.

For systems with finite energy barriers, the transition from glass to
liquid is described by the percolation of bonds between local energy
minima.  Configuration space can be decomposed into basins of
attraction surrounding each local minimum, and every point in
configuration space can be mapped uniquely to a single
basin~\cite{stillingerhiddenstructure}.  At short times the system is
confined to a basin, whereas at long times it will hop from one basin
to another.  Complete structural relaxation occurs once the system's
trajectory percolates configuration space.

To calculate $\tau_\alpha$, it is necessary to specify all possible
percolating bond networks on which the system can relax and then
select the subset that minimizes $\tau_\alpha$.  To build the
ensemble, we prescribe a maximum energy barrier height $n T$ and draw
bonds between minima with barriers below $nT$.  This procedure defines
networks of connected energy minima and if a network percolates it is
included in the ensemble.  For a given $n$, the fraction of bonds
$b(n)$ and the average time $\tau_0(n)$ to make transitions between
two basins are given by
\begin{eqnarray}
b(n) = \int_0^{n T} P_b(E) \,dE ,
\label{bintermsofn}
\\
\tau_0(n) \propto b(n)^{-1} \int_0^{n T} P_b(E) \exp(E/T) \,dE,
\label{tauintermsofn}
\end{eqnarray}
where $P_b(E)$ is the distribution of energy barriers.  For
sufficiently large $N$, we can describe the properties of the percolating
networks using a mean-field description and the only relevant
variable is $b(n)$.
For a given $n$, a percolating network exists if $b(n)$ is larger than a critical value $b_P$, and the alpha relaxation time is
\begin{equation}
\tau_\alpha(n) \propto \Bigg\{ \begin{array}{cc} 
\tau_0(n) \, q^{-2} [b(n)-b_P]^{-2} & \mathrm{for} \, \, q\xi \ll 1, \\
\tau_0(n) \, q^{-6} & \mathrm{for} \, \, q\xi \gg 1.
\end{array}
\label{relaxtimeenergy}
\end{equation}
Near percolation, when relaxation becomes slow, a saddle-point
approximation holds and the system selects the $n^*$ that minimizes
$\tau_\alpha(n)$.  Minimizing Eq.~(\ref{relaxtimeenergy}) for $q \xi \ll 1$ gives
\begin{equation}
\frac{b(n^*)-b_P}{b(n^*)} = \frac{2}{C e^{n^*}/\tau_0(n^*)-1}, 
\label{solveforn}
\end{equation}
where $C$ is the proportionality constant from
Eq.~(\ref{tauintermsofn}) with units of time.  Eq.~(\ref{solveforn})
can be solved to determine $n^*$~\cite{footnote7}.  Since $C e^{n^*} >
\tau_0(n^*)$ for all $T>0$, Eq.~(\ref{solveforn}) predicts that
$b(n^*) > b_P$ and there is no thermodynamic glass transition
at $T>0$ for systems with finite energy barriers.  Furthermore, the
$T$-dependence of $\tau_\alpha(n^*)$ depends on the distribution of
energy barriers.

As in the case of hard spheres, our theory predicts that structural
correlations such as the ISF or spin-spin correlations exhibit
stretched-exponential relaxation with $1/3 \leq \beta \leq 1$. These
limits for the exponent $\beta$ agree with experimental values for
many glass-forming materials~\cite{ngai}.  From our percolation model,
$\beta$ increases slowly with time, but if it is measured near
$\tau_\alpha$ we expect $\beta = 1/3$ for $q \xi \gg 1$ and $\beta=1$
for $q \xi \ll 1$, where $\xi \propto [b(n^*)-b_P]^{-1/2}$.  Our
results for the limiting values of $\beta$ are consistent with
measurements at different $\xi$ values in
Lennard-Jones~\cite{angelani} and magnetic~\cite{ogielski,coey}
glasses and different $q$-values in magnetic~\cite{cbernardi} and
molecular~\cite{cicerone} glasses.

The techniques in this section can be extended to treat hard spheres
beyond the mean-field approximation.  Instead of energy barriers, hard
spheres possess entropic barriers arising from the reduction of the
effective dimension of mobility domains far from CJ states.  For a
system that on average explores mobility domains at a distance $R$
from the CJ state before making a transition to a new mobility domain,
there is an average transition time $\tau_0(R)$ and a corresponding
percolating network characterized by bond fraction $b(R)$.  In the large-$N$
limit, the system will choose the effective domain size $R^*$
that minimizes the relaxation time.

We have introduced a percolation model that gives rise to
stretched-exponential relaxation in glass-forming materials.
Stretched-exponential relaxation is commonly understood by assuming an
underlying heterogeneity, either using trap models with distributions
of waiting times and jump sizes~\cite{trapmodels} or dynamical
heterogeneity in real space~\cite{heterostretched}.  Percolation in
configuration space naturally provides an origin to these
heterogeneities via the structure of the fractal percolating network,
which is organized into densely connected blobs below $\xi$ and
homogeneous nodes above $\xi$~\cite{bloblink}.  Studies of the
geometric properties of mean-field percolating networks, and dynamics
on these networks, will lead to new predictions for slow dynamics and
cooperative motion in glass-forming materials.

Financial support from NSF grant numbers CBET-0348175 (GL,JB) and
DMR-0448838 (GL,CSO), and Yale's Institute for Nanoscience and Quantum
Engineering (GL) is acknowledged.  We also thank the Aspen
Center for Physics where this work was performed.


\begin{thebibliography}{100}
\bibitem{footnote10} We assume each particle is distinguishable, which gives the factor of $N!$.
\bibitem{footnote1} This can be carried out using a procedure in which the system is prepared with particle positions $\{ r_i \}$ at $\phi=0$ and then compressed to $\phi_J$ using the Lubachevsky-Stillinger algorithm~\cite{lubachevsky} in the large compression rate limit.  $\{ r_i \}$ belongs to the basin of attraction of the resultant CJ state.  The set of all basins of attraction tile configuration space and do not overlap.
\bibitem{footnote7}  Eq.~(\ref{solveforn}) also follows trivially if we assume there is a probability $f(E)$ to form bonds between minima with energy barriers $E<n T$, by substituting $P_b(E) \rightarrow f(E)P_b(E)$.
\end{thebibliography}
\end{document}